\documentclass[a4paper,11pt]{article}
\pdfoutput=1 

\usepackage{jinstpub} 

\usepackage{upgreek}
\usepackage{subfigure}
\usepackage{hepnames}
\usepackage{siunitx}

\usepackage{lineno}

\title{Silicon Technologies for the CLIC Vertex Detector}

\author[1]{S. Spannagel\note{Corresponding author.}}
\emailAdd{simon.spannagel@cern.ch}

\abstract{CLIC is a proposed linear $\Ppositron\Pelectron$ collider designed to provide particle collisions at center-of-mass energies of up to \SI{3}{\TeV}.
  Precise measurements of the properties of the top quark and the Higgs boson, as well as searches for Beyond the Standard Model physics require a highly performant CLIC detector.
  In particular the vertex detector must provide a single point resolution of only a few micrometers while not exceeding the envisaged material budget of around 0.2\% X$_0$ per layer.
  Beam-beam interactions and beamstrahlung processes impose an additional requirement on the timestamping capabilities of the vertex detector of about \SI{10}{\ns}.
  These goals can only be met by using novel techniques in the sensor and ASIC design as well as in the detector construction.

  The R\&D program for the CLIC vertex detector explores various technologies in order to meet these demands.
  The feasibility of planar sensors with a thickness of 50--\SI{150}{\um}, including different active edge designs, are evaluated using Timepix3 ASICs.
  First prototypes of the CLICpix readout ASIC, implemented in \SI{65}{\nm} CMOS technology and with a pixel size of $25\times\SI{25}{\um \squared}$, have been produced and tested in particle beams.
  An updated version of the ASIC with a larger pixel matrix and improved precision of the time-over-threshold and time-of-arrival measurements has been submitted.
  Different hybridization concepts have been developed for the interconnection between the sensor and readout ASIC, ranging from small-pitch bump bonding of planar sensors to capacitive coupling of active HV-CMOS sensors.
  Detector simulations based on Geant\,4 and TCAD are compared with experimental results to assess and optimize the performance of the various designs.

  This contribution gives an overview of the R\&D program undertaken for the CLIC vertex detector and presents performance measurements of the prototype detectors currently under investigation.}

\keywords{}

\collaboration[c]{on behalf of the CLICdp collaboration}
\proceeding{Instrumentation for Colliding Beam Physics 2017\\
27 February - 3 March, 2017\\
Budker Institute of Nuclear Physics, Novosibirsk, Russia}

\begin{document}
\maketitle
\flushbottom

\section{Introduction}
\label{sec:intro}

\begin{figure}[tbp]
  \center
  \includegraphics[width=.7\textwidth]{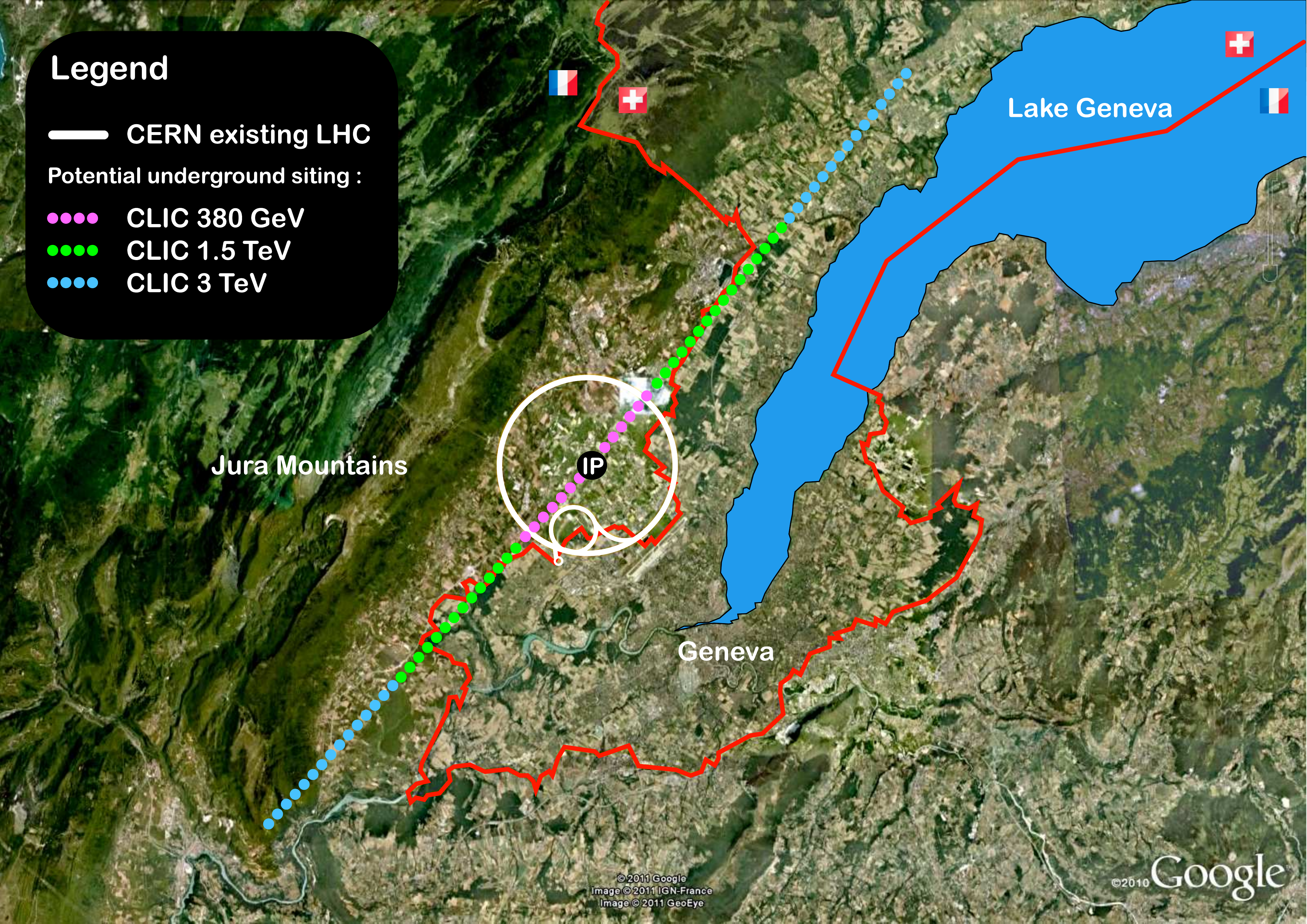}
  \caption[Proposed CLIC layout]{Proposed layout of the CLIC accelerator at CERN. The LHC and SPS rings are indicated and the three CLIC construction stages are shown in different colors.}
  \label{fig:clic}
\end{figure}

The Compact Linear Collider (CLIC)~\cite{clic} is a proposed linear $\Ppositron\Pelectron$ collider based at CERN in Geneva, Switzerland.
Its construction is foreseen in three stages~\cite{clic-baseline} as indicated in Figure~\ref{fig:clic}, with collision energies between \SI{380}{\GeV} and \SI{3}{\TeV}.
CLIC employs a novel two-beam acceleration concept in which a high-current, low-energy drive beam produces the RF for the acceleration of the main beams.
The bunches of the drive beam are accelerated by klystrons at low frequency and a series of delay loops and combiner rings is used to increase the bunch frequency.
High frequency RF power is produced and transferred to the main beams by decelerating the drive beam in normal conducting cavities.
This approach enables RF field gradients of more than \SI{100}{MV/m} to be attained and thus allows a compact collider design.

Owing to this acceleration scheme, the time structure of the CLIC beams has some features which directly impact the design parameters of the detector.
The colliding bunches are arranged in bunch trains with a total length of \SI{156}{\ns} and a repetition rate of \SI{50}{\Hz}.
One train consists of 312 bunches that are spaced by \SI{0.5}{\ns}.
The low train repetition rate allows parts of the front-end electronics to be switched off between the bunch trains, the so-called \emph{power pulsing} scheme, which significantly reduces the average power consumption of the detector and hence the required cooling infrastructure.

In order to reach the anticipated luminosity of about $\mathcal{L} = \SI{6e34}{\per \square \cm \per \s}$, each bunch of \num{1e9} particles is compressed to a size of about $\sigma_x \times \sigma_y \times \sigma_z \approx \SI{40}{\nm} \times \SI{1}{\nm} \times \SI{44}{\um}$.
The high particle density in the bunch and the very small extent lead to strong electromagnetic interactions between the colliding bunches and particles within the bunch.
The resulting beamstrahlung~\cite{ep-limits} reduces the total collision energy of the particles and produces a large amount of background particles, especially in the forward regions of the experiment.
The impact of this beam-induced background can be reduced by applying timing cuts on the detector signals to select the hard interaction of interest.
For most sub-detectors, a time resolution of about \SI{10}{\ns} is foreseen; a correspondingly precise time stamping capability is required for the tracking systems of the CLIC detector.
The CLIC vertex detector is the innermost component and provides crucial information for the identification of secondary decay vertices as well as for the reconstruction of particle trajectories.

The current proposal for the vertex detector comprises six barrel layers of silicon pixel detectors, arranged in three double layers with modules on both sides of the mechanical support structure.
This approach allows the material budget, which is currently envisaged to be around 0.4\%\,$X_0$ per double layer, to be minimized.
The forward regions are also covered by three double layers, which are mounted in a spiral geometry to allow air cooling as described below.
The current detector design foresees a hybrid pixel detector with a thin readout ASIC and silicon sensor, each having a thickness of about \SI{50}{\um}.
The pixel size is $25\times\SI{25}{\um \squared}$ and both planar sensors and capacitively coupled active sensors are considered.
A single hit resolution of about \SI{3}{\um} is required to reach the desired impact parameter resolution for flavor tagging.

By power pulsing the front-end ASICs, the mean power consumption of the vertex detector is limited to below \SI{50}{\milli \watt \per \square \cm}.
The low power consumption and the detector geometry allow cooling using forced air flow which further reduces the overall material budget by not requiring cooling pipes and coolant.
The spiral geometry of the forward regions acts as an airflow guide and ensures a proper distribution of the air in the detector.
This approach has been shown to perform well in previous experiments featuring a barrel detector only~\cite{starpxl} as well as in a mechanical mockup of the CLIC vertex detector~\cite{airflow}.

\section{The CLICpix Readout Chip Prototypes}
\label{sec:clicpix}

First prototype readout ASICs following the requirements of the CLIC vertex detector have been developed.
The CLICpix ASIC design~\cite{clicpix} is derived from the Timepix/Medipix chip family~\cite{timepix,timepix3,medipix} and is implemented in a commercial \SI{65}{\nm} CMOS process with an active pixel array of $64\times 64$ pixels and a pitch of \SI{25}{\um}.
Each pixel features a charge sensitive amplifier (CSA), a discriminator as well as digital logic for the simultaneous measurement of a 4-bit time-of-arrival (ToA) and a 4-bit time-over-threshold (ToT) for every hit.
The chip is operated in a shutter-based readout mode which is well-suited for the CLIC beam structure.
Furthermore, it supports power pulsing of the analog circuitry and features an optional on-chip data compression.
The chip has been extensively tested both in laboratory measurements and test beam campaigns and has shown to perform well as will be discussed later.


The CLICpix2 ASIC represents an advancement of the CLICpix design and is produced in the same \SI{65}{\nm} CMOS process. 
The chip features a larger active matrix with $128 \times 128$ pixels, which corresponds to a total active area of about $3.2\times\SI{3.2}{\square \mm}$.
The pixel circuitry provides measurements of the ToA and ToT with 8-bit and 5-bit precision, respectively.
Relative to the first prototype, the new ASIC has an improved noise isolation and a source of cross talk observed in CLICpix has been identified and removed.
Furthermore, the new chip contains an integrated circuit for test pulses as well as a band gap reference and the capability of power pulsing for both the analog and the digital parts.
A first production of the CLICpix2 has been received and is currently being tested in the laboratory. 

\section{Planar Silicon Sensor Studies}
\label{sec:planar}

Planar silicon sensors are one of the options considered for the CLIC vertex detector.
The following sections provide a brief overview of the different designs and assemblies under investigation.

\subsection{Sensor Studies using Timepix ASICs}

Different thin planar sensors with slim-edge and active-edge design have been tested using the Timepix~\cite{timepix} and Timepix3~\cite{timepix3} ASICs for the readout.
Assemblies with \SI{50}{\um} to \SI{500}{\um} thick sensors have been produced and tested.
The thinnest full assembly tested is a \SI{100}{\um} thin sensor bump bonded to a Timepix chip thinned down to \SI{100}{\um}~\cite{thin-timepix}.

The active edge technology~\cite{edgeless} is an interesting option to reduce the amount of passive sensor area and thus the overall material budget  of the detector.
For this, the backside implant is extended to the edge of the sensor.
The deep reactive ion etching (DRIE) process is used in order to achieve a smooth edge which allows for the implantation.
Extending the backside implant alters the electric field configuration and allows the distance between the last pixel implant and the physical edge of the sensor to be reduced substantially.

Different sensor edge designs are possible, including the use of guard rings to shape the electric field.
Several sensor designs with different guard ring configurations have been produced by Advacam~\cite{advacam} and tested for their charge collection behavior and efficiency.
All sensors feature a pitch of \SI{55}{\um} in order to facilitate testing with the Timepix3~\cite{timepix3} readout chip.

\begin{figure}[tbp]
  \center
  \includegraphics[width=.33\textwidth]{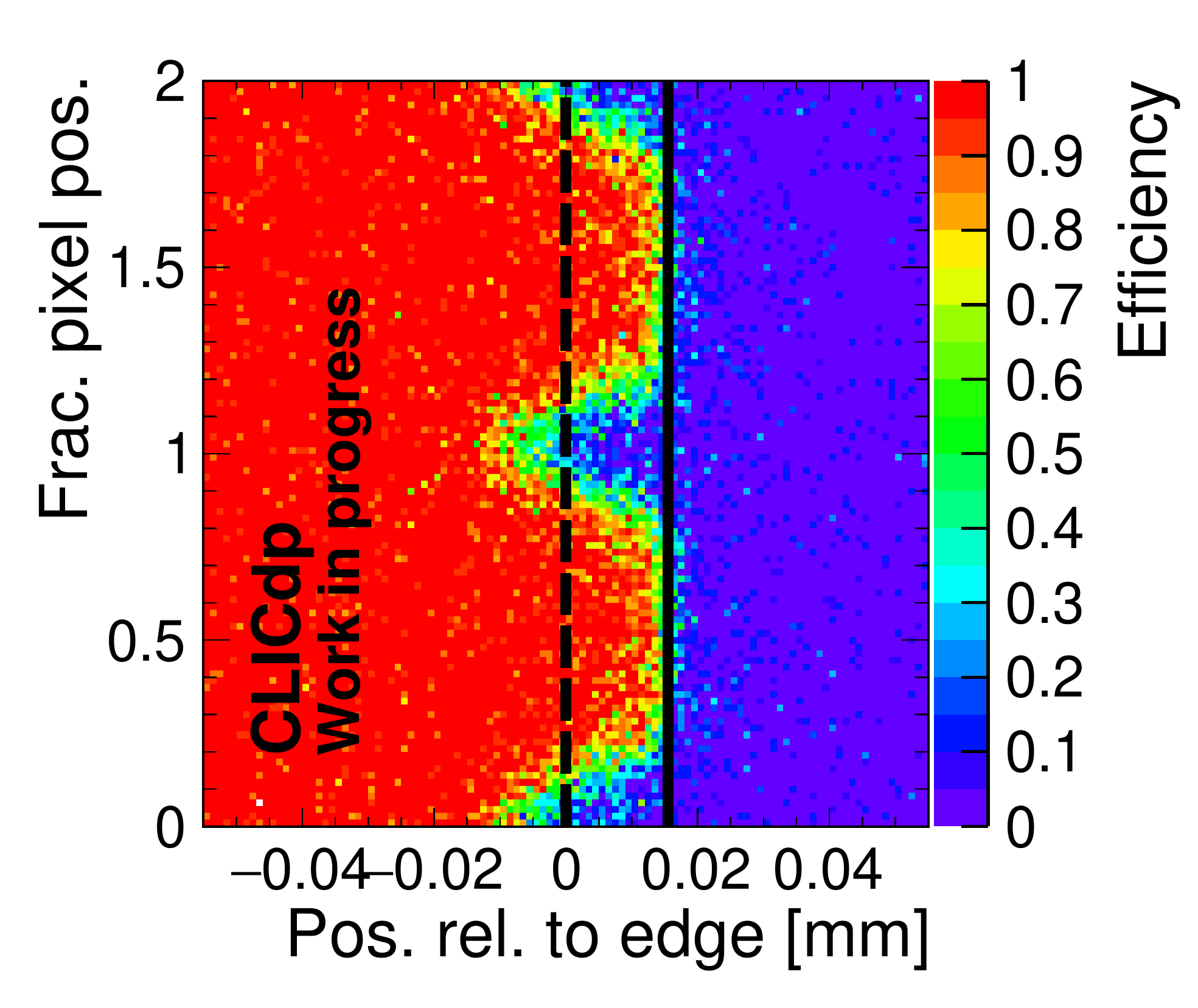}%
  \includegraphics[width=.33\textwidth]{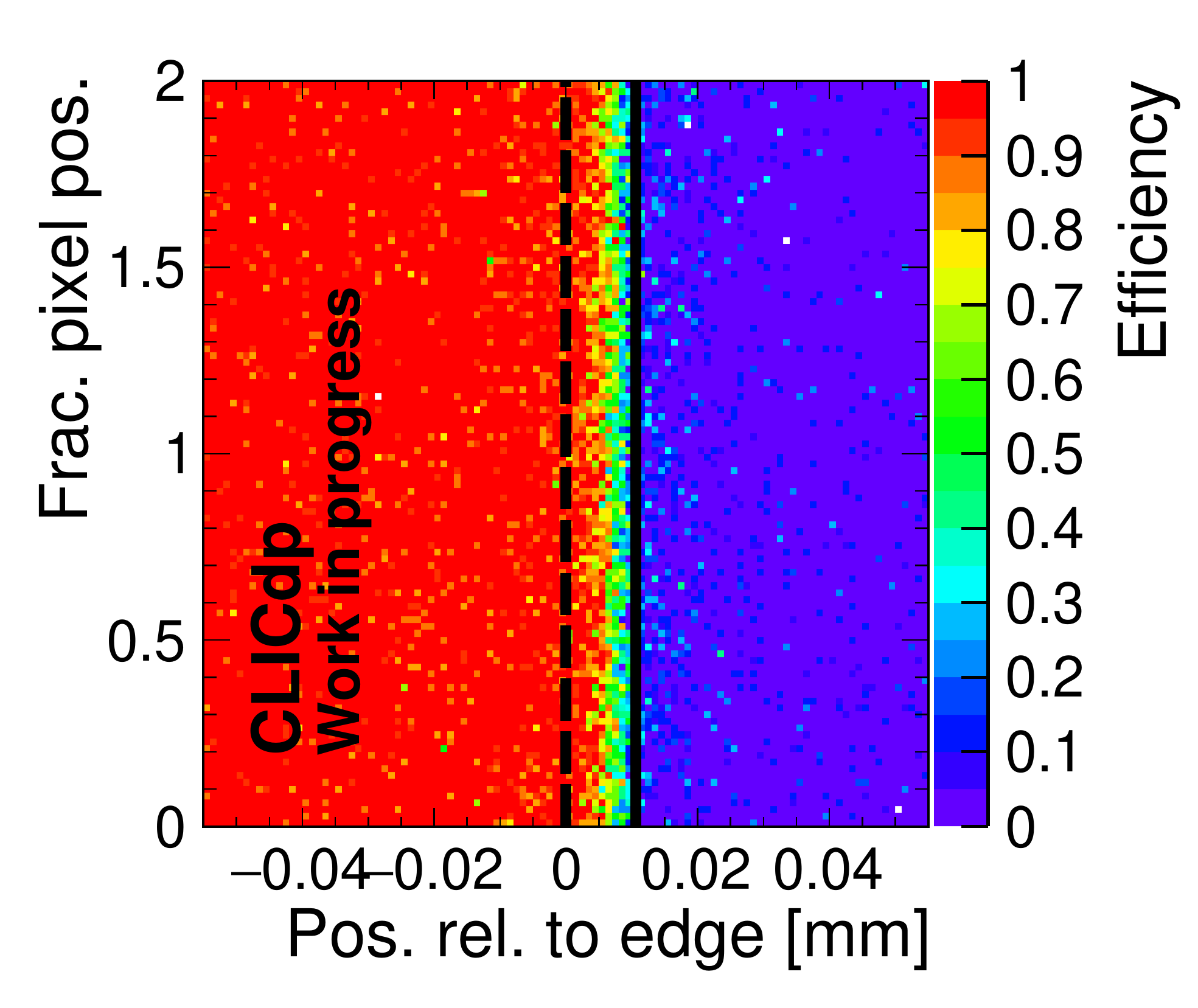}%
  \includegraphics[width=.33\textwidth]{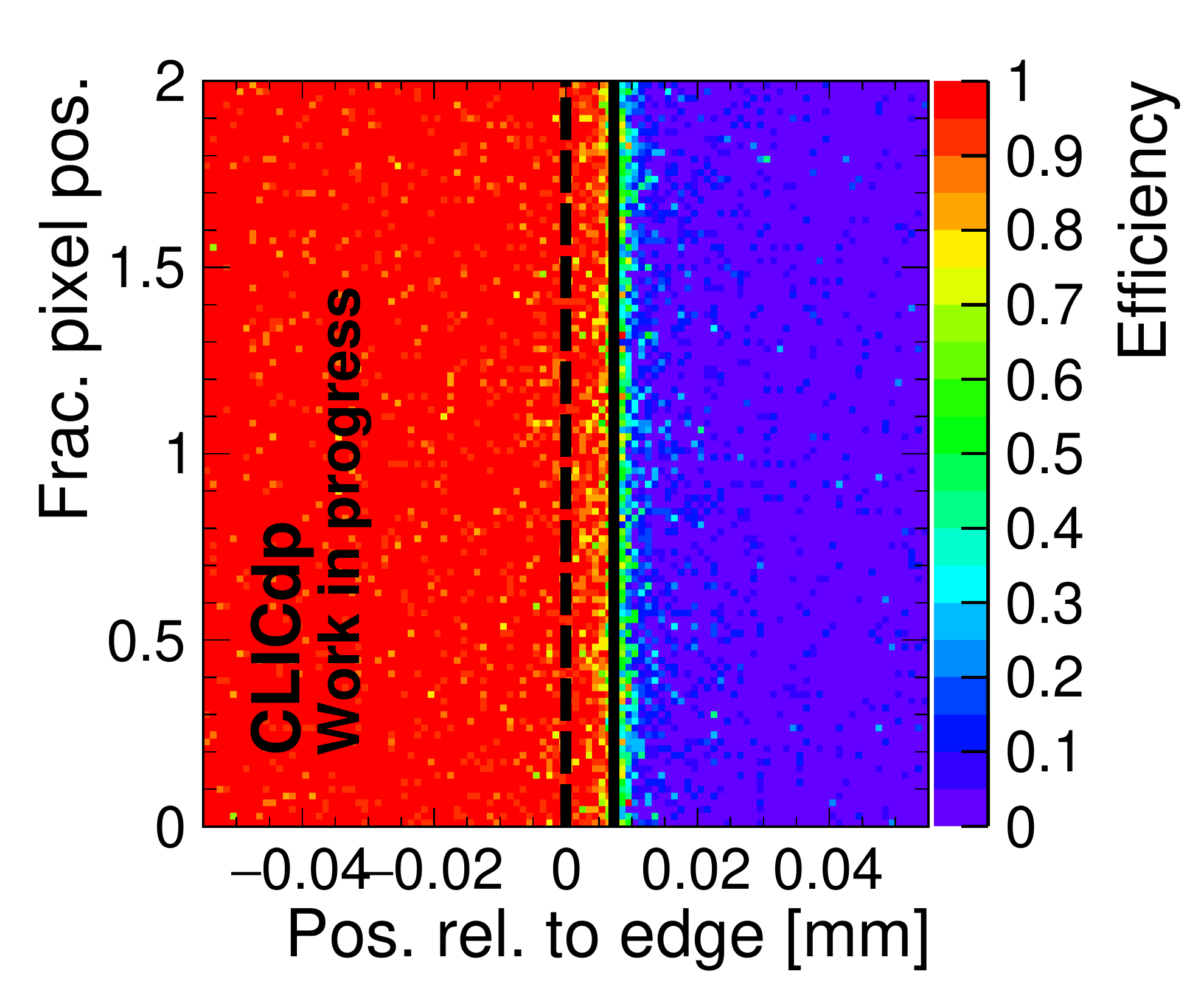}
  \caption[Single hit efficiency of active edge sensors]{Single hit efficiency for active edge sensors with grounded guard rings~(left), floating guard rings~(center) and without guard rings~(right). The distance from the pixel implant to the sensor edge is \SI{28}{\um}, \SI{23}{\um} and \SI{20}{\um}, respectively. The dashed lines indicate the implant end while the solid lines represent the physical edge of the sensor. Modified from~\cite{nilou-proc}.}
  \label{fig:active}
\end{figure}

The first design tested includes guard rings which are kept at ground potential by connecting them to the readout ASIC with an additional row of bump bonds.
Figure~\ref{fig:active}~(left) shows the single hit efficiency for the sensor edge.
Here, all edge pixels have been folded into two pixel cells in order to increase statistics.
It can be seen that the grounded guard rings lead to a sizable inefficiency near the edge in the regions between pixel implants as they compete with the guard rings for charge collection in particular between the pixels, where the distance to the guard ring might be smaller than to the adjacent pixels.

\begin{figure}[tbp]
  \center
  \includegraphics[width=.33\textwidth]{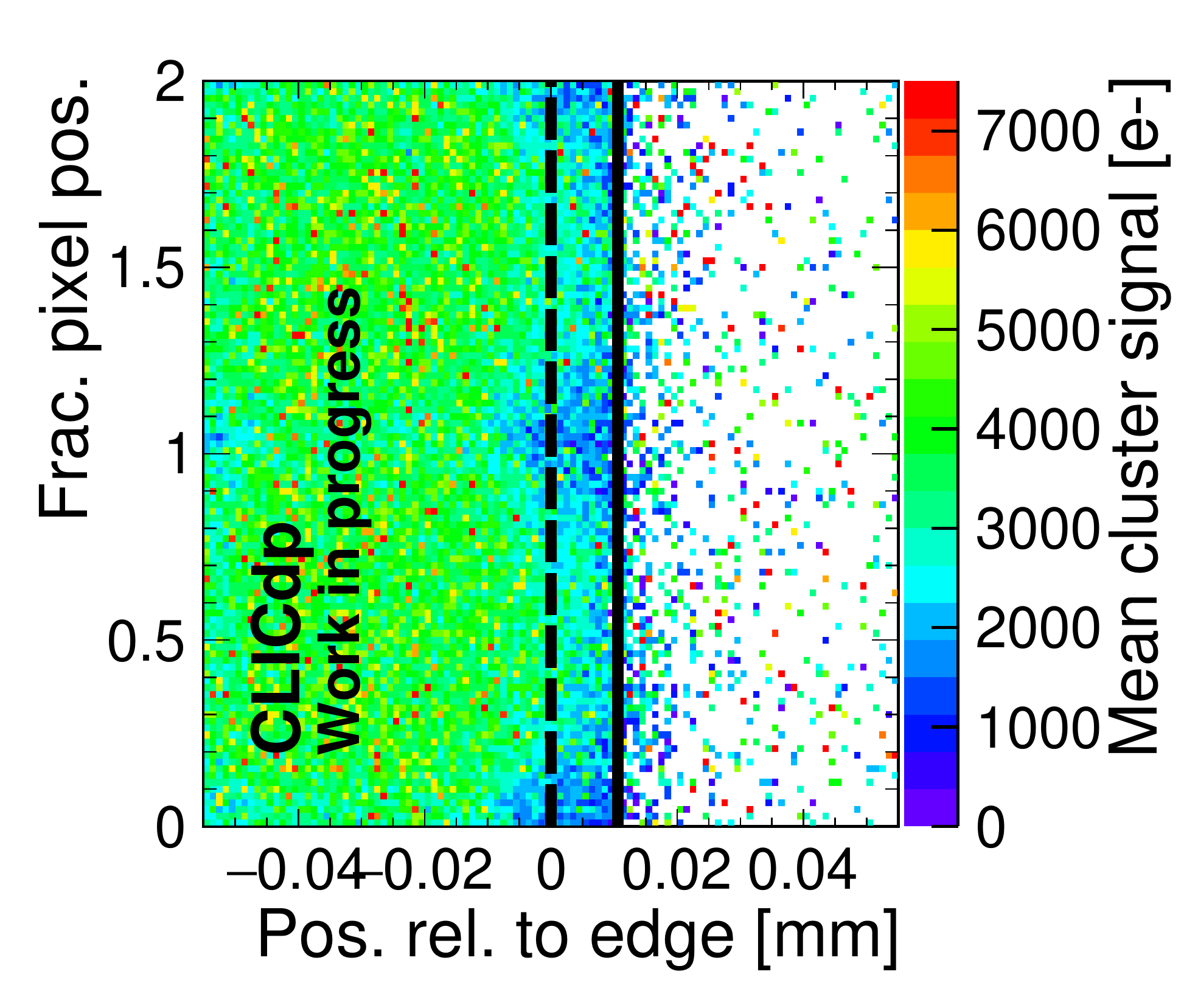}%
  \includegraphics[width=.33\textwidth]{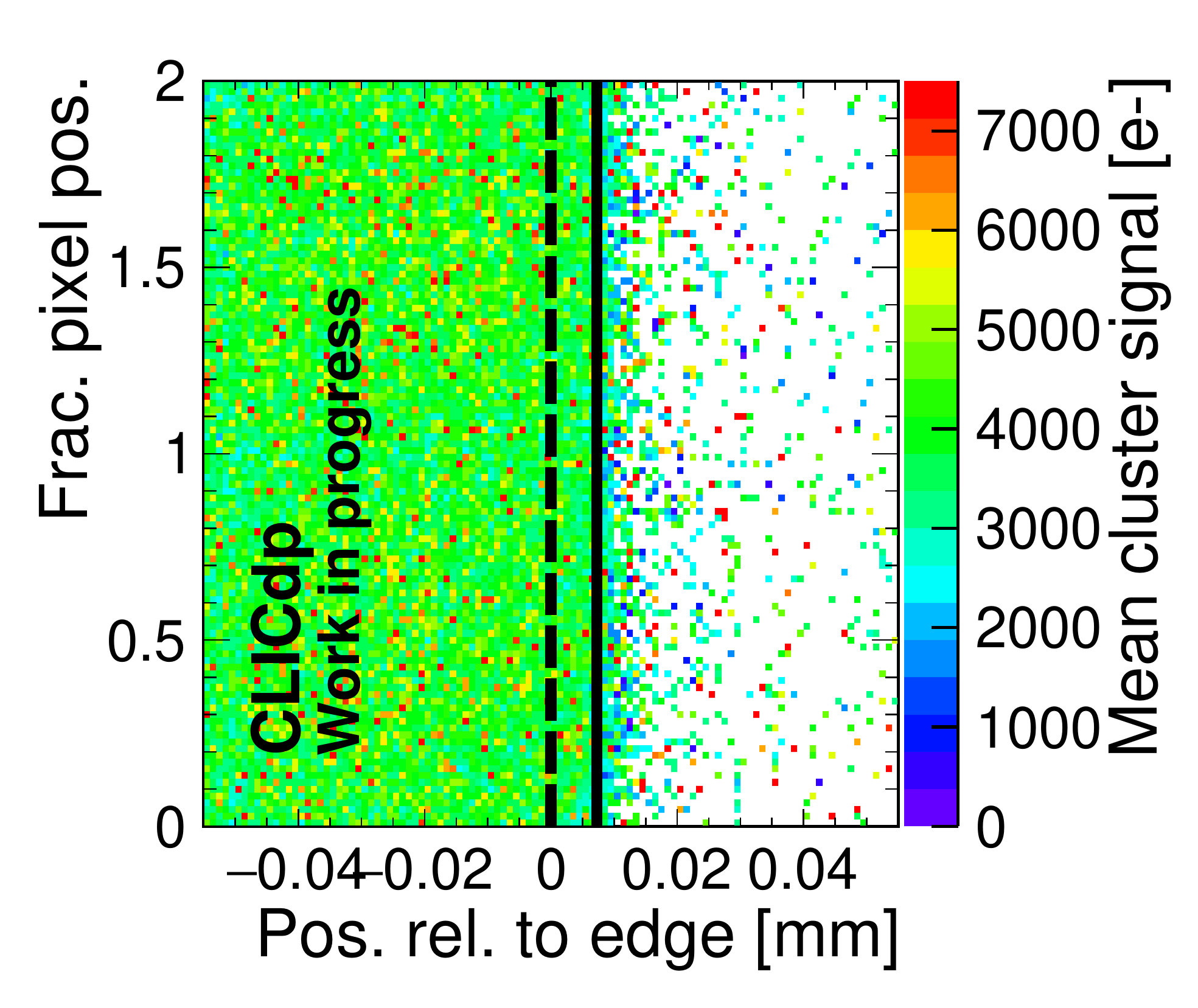}
  \caption[Mean cluster charge of active edge sensors]{Mean cluster charge of active edge sensors with floating guard rings~(left) and without guard rings~(right). The distance from the pixel implant to the sensor edge is \SI{23}{\um} and \SI{20}{\um}, respectively. The dashed lines indicate the implant end while the solid lines represent the physical edge of the sensor.}
  \label{fig:activeq}
\end{figure}

In the second design, the guard rings are left unconnected and thus floating.
The impact of this change is demonstrated in Figure~\ref{fig:active}~(center), where only a marginal drop of the efficiency can be observed towards the physical cut edge of the sensor.
Figure~\ref{fig:activeq}~(left) shows the mean cluster signal over a two-pixel region of the same sensor, and charge losses of tracks close to the sensor edge due to charge being collected on the floating guard ring are visible.

Full efficiency up to the physical edge of the sensor could only be achieved with the third design without guard rings.
Figure~\ref{fig:active}~(right) shows that a uniform and very high single hit efficiency can be achieved.
The previously discussed charge losses close to the edge are mitigated in this design as can be seen in Figure~\ref{fig:activeq}~(right).

These findings have also been studied in simulation using TCAD and good agreement is found.
All sensors could be operated well above their depletion voltage without breakdown.

\subsection{Sensor Studies with the CLICpix ASIC}

Thin planar sensors with a pixel pitch of \SI{25}{\um} have been tested using the CLICpix ASIC.
The ASICs have been produced on a multi-project wafer and the chips could thus not be post-processed at wafer level.
While large-scale bump deposition on sensors with \SI{25}{\um} pitch is established at wafer level~\cite{bumpbonds-moench}, a custom process had to be developed at the Stanford Linear Accelerator laboratory (SLAC)~\cite{bumpbonding} in order to allow bump bonding of single chips and sensors after dicing.
For this, indium bumps were deposited on both ASIC and sensor prior to the flip chip process.
Because of the small chip size and pitch, a large spread in quality between different assemblies has been observed.
The optimization of the bump bonding process is ongoing and some of the observed issues have been understood in the meantime.
Nonetheless, samples produced with several \SI{200}{\um}-thick slim-edge planar sensor and a \SI{50}{\um}-thick active edge sensor allowed measurements and have been characterized.

\begin{figure}[tbp]
  \center
  \includegraphics[width=.45\textwidth]{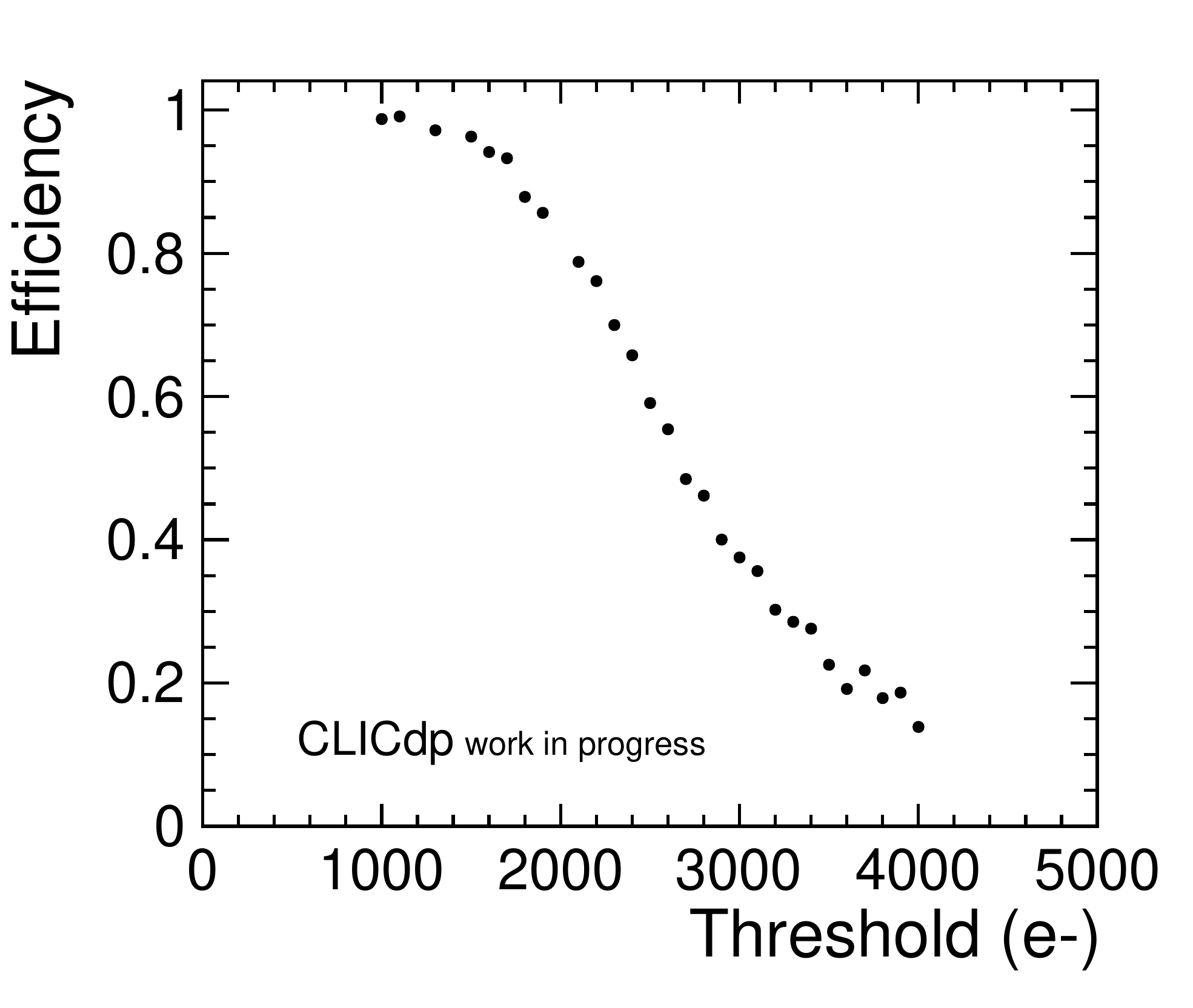}%
  \includegraphics[width=.45\textwidth]{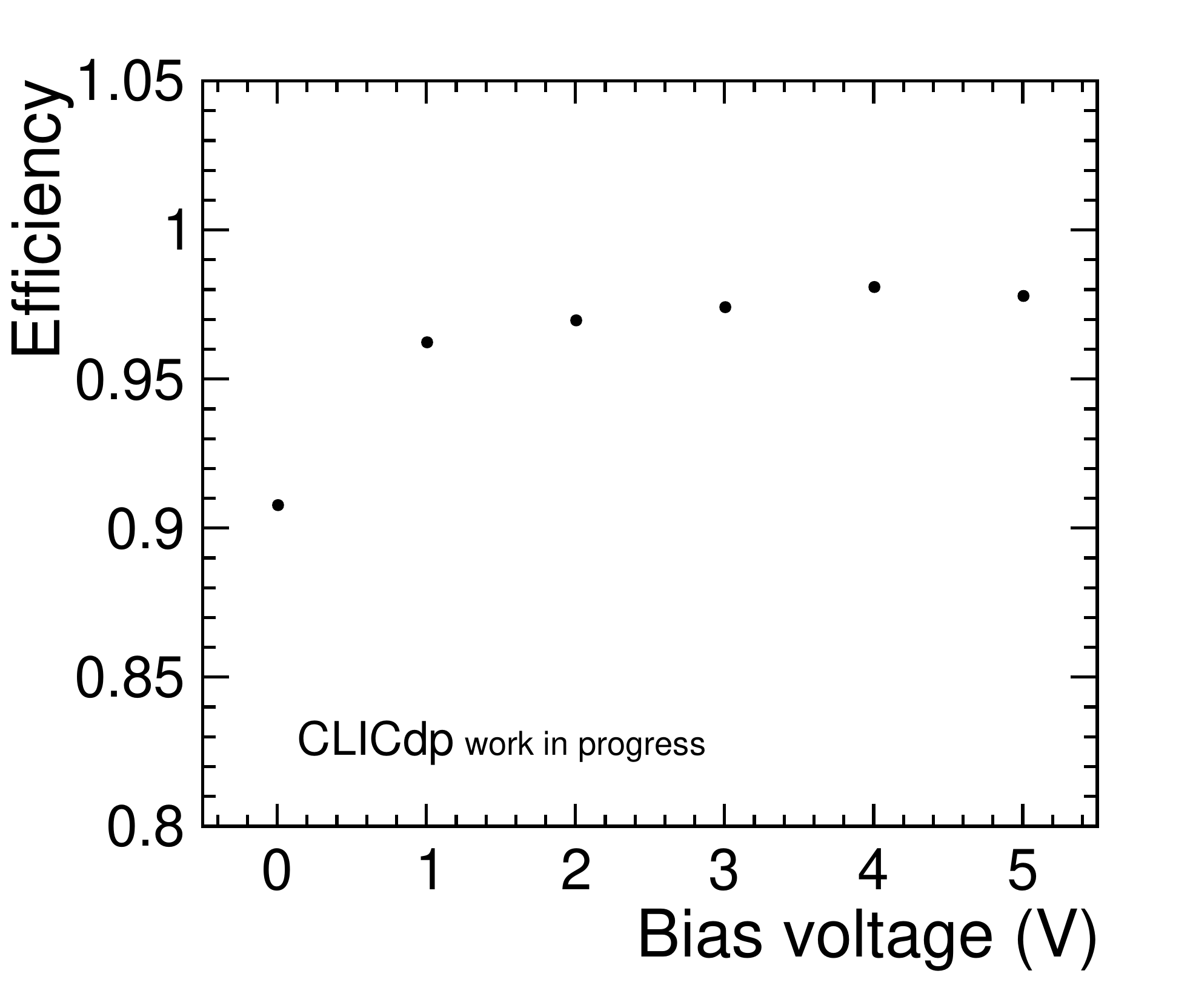}
  \caption[Efficiency as a function of threshold and bias voltage]{Efficiency of a \SI{50}{\um} thick planar silicon sensor bump bonded to the CLICpix ASIC as a function of the applied charge threshold~(left) and the bias voltage at a threshold of \SI{1.2}{ke}~(right).}
  \label{fig:smallpitch}
\end{figure}

Figure~\ref{fig:smallpitch}~(left) shows the efficiency of the \SI{50}{\um} sensor assembly as a function of the charge threshold applied, with an efficiency of around 99\% at a threshold of \SI{1}{ke}.
With lower charge thresholds, achievable with the design optimizations implemented in the CLICpix2 prototype, even higher efficiencies should be feasible.
Figure~\ref{fig:smallpitch}~(right) shows the efficiency of the same assembly as a function of the applied bias voltage at a fixed charge threshold of \SI{1.2}{ke}.
While full depletion is reached at around \SI{2}{\V}, the detector is observed to be very efficient even in the absence of external biasing due to the built-in potential and an offset of the pixel ground level resulting from the biasing of the CMOS circuitry.

\section{Capacitively Coupled Active Sensors}
\label{sec:capacity}

Active sensors implement some functionality for charge collection and amplification directly in the sensor.
However, with standard CMOS processes, the charge is mostly collected by diffusion as the sensor cannot be fully depleted.
High Voltage (HV) CMOS processes~\cite{hvcmos} allow transistors to be shielded by an additional deep $n$-well and thus enable a fast signal collection in the sensor by applying a moderate bias voltage.
This removes the need for bump bonding as the sensor can be directly glued to the readout ASIC and the amplified signal transferred via capacitive coupling.

For the CLIC vertex detector, a Capacitively Coupled Pixel Detector (CCPDv3)~\cite{hvcmos2} has been designed. 
It is produced in a commercial \SI{180}{\nm} HV-CMOS process and features the same footprint as the CLICpix ASIC with an active pixel matrix of \num{64x64} pixels with a pitch of \SI{25}{\um}.
Each pixel contains a CSA as well as a second inverting amplifier stage.
The chip contains only limited standalone readout capabilities and is designed for being read out by the CLICpix ASIC.
Although comprehensive test beam studies~\cite{capacitively} have shown good performance of the sensor, some parameters, e.g., the amplifier peaking time of around \SI{120}{\ns}, do not yet meet the requirements for the CLIC vertex detector.
Studies of the impact of misalignment of the two components during fabrication on the total coupling capacitance and the cross coupling between pixels have been performed~\cite{daniel-vertex}.

A major redesign of this sensor has been developed and produced, called CLIC Capacitively Coupled Pixel Detector (C3PD), including power pulsing circuitry, an improved test pulse injection and a single amplifier stage.
Standalone measurements in the laboratory show an improved amplifier rise time of about \SI{20}{\ns} and high average gain of \SI{190}{\milli \V \per ke}.
Some samples have been thinned down to \SI{50}{\um} and no performance degradation has been observed.
A full test of the sensor using the CLICpix2 ASIC is foreseen for the coming test beam campaigns.

\section{Summary and Outlook}
\label{sec:summary}

The CLIC accelerator technology and the resulting beam structure impose stringent requirements on the CLIC vertex detector.
New readout ASICs and sensor technologies are being developed to meet the goals of a low mass detector with high spatial and temporal resolution.

The first readout chip prototype, CLICpix, performed well in several test beam campaigns.
CLICpix2 represents an advancement of this design with several improvements and a larger active area, and is currently under investigation in the laboratory.

Several technologies are considered for planar silicon sensors.
First samples of sensors thinned to \SI{50}{\um} have been tested and perform well.
Active edge technology would allow the reduction of passive material by extending the active region of the sensor to the physical cutting edge.
Different guard ring designs have been investigated and sensors without guard rings have been found to provide the best efficiency up to the sensor edge.
The small pitch of \SI{25}{\um} poses a challenge for bump bonding but processes have been developed to allow bump bonding at single chip level.

The concept of capacitively coupling an active sensor to the readout ASIC is under investigation as a possible alternative to standard planar silicon sensors.
Assemblies of the capacitively coupled active sensor CCPDv3 and the CLICpix ASIC have been produced and successfully operated.
Based on this experience, an advancement of the active sensor, C3PD, has been designed.
The chip has shown to perform well in laboratory measurements and will be tested in a particle beam using the CLICPix2 chip for the readout.

\section*{Acknowledgment}

This project has received funding from the European Union's Horizon 2020 research and innovation programme under grant agreement No 654168.

\bibliographystyle{unsrt}

\bibliography{bibliography}
\end{document}